\documentclass[11pt]{article}
\usepackage[utf8]{inputenc}
\usepackage{graphicx}
\usepackage{subcaption}
\usepackage{booktabs}
\usepackage{hhline}
\usepackage{amssymb}
\usepackage{amsmath}
\usepackage{enumitem}
\usepackage{graphics}
\usepackage{setspace}
\usepackage{sparklines}
\usepackage{colortbl}
\usepackage{xcolor}
\usepackage{diagbox}
\usepackage[a4paper, inner=2.54cm,outer=2.54cm,bottom=2.54cm,top = 2.54cm]{geometry}
\usepackage{slashbox}
\usepackage{sectsty,textcase}
\sectionfont{\MakeTextUppercase}
\makeatletter

\usepackage[
backend=biber,
style=apa,
isbn=false,
doi=false,
sorting=nyt
]{biblatex}
\providecommand{\keywords}[1]
{
  \small	
  \textbf{\textit{Keywords--}} #1
}
\title{Testing parametric additive time-varying GARCH models}
\author{Niklas Ahlgren$^1$, Alexander Back$^1$\thanks{Alexander Back gratefully acknowledges financial support from the OP Research Foundation (grants No. 20220141, 20210112 and 20200213). Material from this paper has been presented at the 16th International Conference on Computational and Financial Econometrics (CFE 2022), London, December 2022. Comments from participants are gratefully acknowledged. We wish to thank Yongmiao Hong and Esther Ruiz for many thoughtful comments.} \& Timo Teräsvirta$^{2}$}
\date{%
    $^1$Hanken School of Economics\\
    $^2$CoRE, Aarhus BSS, Aarhus University\\
    [2ex]
}

\begin{document}

\maketitle
\thispagestyle{empty}
\begin{abstract}
\footnotesize
We develop misspecification tests for building additive time-varying (ATV-)GARCH models. In the model, the volatility equation of the GARCH model is augmented by a deterministic time-varying intercept modeled as a linear combination of logistic transition functions. The intercept is specified by a sequence of tests, moving from specific to general. The first test is the test of the standard stationary GARCH model against an ATV-GARCH model with one transition. The alternative model is unidentified under the null hypothesis, which makes the usual LM test invalid. To overcome this problem, we use the standard method of approximating the transition function by a Taylor expansion around the null hypothesis. Testing proceeds until the first non-rejection. We investigate the small-sample properties of the tests in a comprehensive simulation study. An application to the VIX index indicates that the volatility of the index is not constant over time but begins a slow increase around the 2007–2008 financial crisis.
\end{abstract}
\keywords{
Misspecification testing, nonstationarity, nonlinear GARCH, smooth transition, time-varying intercept, VIX
}

\newpage
\section{Introduction}
\setcounter{page}{1}

Commonly used volatility models like the ARCH model of \textcite{Engle82} and the generalized ARCH (GARCH) model of \textcite{Bollerslev86} and \textcite[pp. 78--79]{Taylor86} assume that volatility is stationary. This assumption is a reasonable one to make when the time series is relatively short. However, the volatility of an asset or index may change over time. This was recognized by \textcite{mikoschstarica2004a} and \textcite{staricagranger2005} who attempt to identify intervals of homogeneity, which they define as intervals where a stationary model describes the data well. \textcite{Hardle2003} used local homogeneity for joint modeling of variances and covariances. 

In long time series, gradual structural change makes the assumption of stationarity untenable. To model this type of smoothly changing, nonstationary volatility, in a parsimonious way, \textcite{abt2023est}, ABT for short, proposed the additive time-varying (ATV-)GARCH model with a time-varying intercept. The model nests a standard stationary GARCH model but is not identified when the data are generated by the latter. Consistent estimation requires time-variation in the intercept. The purpose of the present work is to derive appropriate test statistics for testing stationary GARCH against ATV-GARCH. More generally, the tests may be used for determining the parametric structure of the ATV-GARCH model.

Several researchers have proposed volatility models with time-varying parameters. \textcite{DahlhausSR06} generalized the stationary ARCH model to time-varying ARCH. \textcite{rao2006} introduced the time-varying GARCH model. For further developments, see \textcite{Rohan13}, \textcite{chenhong} and \textcite{kristensen2019local}. These models are nonparametric. \textcite{Truquet17} introduced a semiparametric ARCH model where some, but not all, parameters are time-varying. In statistical tests he finds that the null hypothesis of parameter constancy tends to be rejected for the intercept, but not for the ARCH parameters. 

Another common approach to capture the nonstationarity in volatility is to decompose the variance multiplicatively into a slow-moving and a transient component. \textcite{Feng04} and \textcite{BellegemS04} were the first to propose such a decomposition. Their approach is nonparametric. \textcite{Engle&Rangel08} proposed a decomposition where the slow-moving component consists of exponential quadratic splines. For surveys of the earlier literature, see \textcite{VBellegem12} and
\textcite{AmadoST19}. \textcite{AmadoT13} suggested a multiplicative time-varying GARCH model where the slow-moving component is given by a linear combination of logistic transition functions in rescaled time. For another recent model with a multiplicative structure, see \textcite{ES2018}.

The paper is structured as follows. In Section 2, we briefly introduce the ATV-GARCH model and derive LM-type tests for additive misspecification. In Section 3, we investigate the small-sample properties of the tests in a simulation study featuring types of nonlinearity in the transition function not considered in previous work. In Section 4, we apply the tests to daily returns of the implied volatility index VIX. Section 5 concludes.

\section{Test for additive misspecification}
Various misspecification tests have been derived for ARCH and GARCH models. \textcite{Engle82} constructed a test for ARCH errors. \textcite{Bollerslev86} derived LM statistics for testing the standard GARCH model against various types of misspecification. Subsequently, several authors have proposed misspecification tests for GARCH models. For example, see \textcite{EngleNg93} who derive an LM test for asymmetric impact of news on volatility.

In this section, we introduce the ATV-GARCH model and use the LM approach to derive tests for additive misspecification of the intercept. 

\subsection{The ATV-GARCH model}
The ATV-GARCH model augments the volatility equation of the GARCH model by a deterministic time-varying intercept:
\begin{equation}
	X_{t}=\sigma _{t}Z_{t}, \ t=1,\ldots,T,
	\label{eq:GARCH}
\end{equation}
where $X_{t}$ is the daily log-return of a financial asset, $Z_{t}$ is IID$(0,1)$ and
\begin{equation}
	\sigma_{t}^{2}= \alpha_{0}+g(t/T;\boldsymbol{\theta}_1)
    +\sum_{i=1}^{p}\alpha_{i}X_{t-i}^{2}+\sum_{j=1}^{q}\beta_{j}\sigma_{t-j}^{2}.\label{eq:themodel}
\end{equation}
The GARCH parameters satisfy the restrictions  $\alpha_0>0$, $\alpha_1, \alpha_2, \ldots, \alpha_p>0$, $\beta_1, \beta_2, \ldots, \beta_q>0$. The time-varying intercept satisfies the positivity restriction $\alpha_0 + g(t/T;\boldsymbol{\theta}_1 )>0$. The function $g(\cdot)$ is parameterized as a linear combination of logistic transition functions in rescaled time,
\begin{equation}
	g_t = g(t/T;\boldsymbol{\theta}_1 ):=\sum_{l=1}^{L}\alpha_{0l}G_{l}\left(t/T;\gamma_{l},c_{l}\right),\label{eq:tvintercept}
\end{equation}
where
\begin{equation}
	G_l(t/T;\gamma_l,c_l)=\left(1+\exp\left\{ -\gamma_l\left(t/T-c_l\right)\right\} \right)^{-1} \label{eq:lfunct}
\end{equation}
is the generalized logistic function. The parameters satisfy $\gamma_{l}>0,\ c_{1}<c_{2}<\ldots<c_{L}, l=1,\ldots,L$.
Since $g_t$ is Lipschitz continuous and bounded, the ATV-GARCH process is locally stationary (see \cite{rao2006}). For asymptotic inference, the process needs to have a finite fourth moment, which requires the assumption $\mathbb{E}|Z_t|^{4+\delta}<\infty$ for some $\delta>0,$ as well as restrictions on the GARCH parameters equal to the ones in \textcite{he1999} and \textcite{LingMcaleer}.

As ABT showed, the ATV-GARCH model (\ref{eq:GARCH})--(\ref{eq:lfunct}) bears similarity to the multiplicative time-varying (MTV-)GARCH model of \textcite{AmadoT13}. It is defined as 
\begin{equation*}
    \sigma_t^2 = h_tg_t, \quad h_t = \alpha_0 + \alpha_1X^2_{t-1}/g_{t-1} + \beta_1h_{t-1},
\end{equation*}
where $g_t$ is defined as in (\ref{eq:tvintercept}). Substituting the expressions for $h_t$ and $g_t$ into the multiplicative decomposition of $\sigma^2_t$ yields
\begin{equation} \sigma^2_t =\alpha_0g_t + \alpha_1\frac{X^2_{t-1}g_t}{g_{t-1}} + \beta_1\frac{\sigma^2_{t-1}g_t}{g_{t-1}} \approx \alpha_0g_t + \alpha_1X^2_{t-1} + \beta_1 \sigma^2_{t-1}. \label{eq:lipapprox}
\end{equation}
The approximation (\ref{eq:lipapprox}) follows from Lipschitz continuity of $G(\cdot)$; $g_t/g_{t-1} \approx 1$ when $T$ is large. The MTV-GARCH model asymptotically has a representation as an ATV-GARCH model. \textcite{Truquet17} made a similar argument for ARCH. 

ABT consider estimation of the ATV-GARCH model (\ref{eq:GARCH})--(\ref{eq:lfunct}) by quasi maximum likelihood (QML), whereas the focus of this article is on testing for misspecification. The Gaussian log-likelihood function of $(X_{1} \ldots, X_{T})$ is given by $L_{T}(\boldsymbol{\theta}) = (1/T)\sum^T_{t=1}l_{t}(\boldsymbol{\theta})$, where
\begin{equation*}
	l_{t}(\boldsymbol{\theta})=-\frac{1}{2}\left[\log \sigma^{2}_{t}(\boldsymbol{\theta})+\frac{X^2_{t}}{\sigma^{2}_{t}(\boldsymbol{\theta})}\right]
\end{equation*}
is the log-likelihood for observation $t$. The QMLE is defined as $\widehat{\boldsymbol{\theta}}_T=\arg \underset{\boldsymbol{\theta}\in\Theta}{\max} \ L_T(\boldsymbol{\theta})$. The score and Hessian evaluated at the "true" parameter vector $\boldsymbol{\theta}_0$ are given by (omitting constants)
$\mathbf{s}_{T}(\boldsymbol{\theta}_0)=(1/T)\sum_{t=1}^{T} \mathbf{s}_t(\boldsymbol{\theta}_0)$
and $\mathbf{H}_{T}(\boldsymbol{\theta}_0)=(1/T)\sum_{t=1}^{T} \mathbf{H}_t(\boldsymbol{\theta}_0)$, respectively, where
\begin{align}
\mathbf{s}_t(\boldsymbol{\theta}_0)&= \left(1-Z^2_t\right)\frac{1}{\sigma^2_t(\boldsymbol{\theta}_0)}\frac{\partial \sigma^2_t(\boldsymbol{\theta}_0)}{\partial\boldsymbol{\theta}},\label{eq:score}
\end{align}
is the score and

\begin{align}
\boldsymbol{H}_t(\boldsymbol{\theta}_0)&=\left(1-Z^2_t\right)\frac{1}{\sigma^2_t(\boldsymbol{\theta}_0)}\frac{\partial^2 \sigma^2_t(\boldsymbol{\theta}_0)}{\partial\boldsymbol{\theta}\partial\boldsymbol{\theta}^{\intercal}}+\left(2Z^2_t-1\right)\frac{1}{\sigma^2_t(\boldsymbol{\theta}_0)}\frac{\partial \sigma^2_t(\boldsymbol{\theta}_0)}{\partial\boldsymbol{\theta}}\frac{1}{\sigma^2_t(\boldsymbol{\theta}_0)}\frac{\partial \sigma^2_t(\boldsymbol{\theta}_0)}{\partial\boldsymbol{\theta}^{\intercal}}\label{eq:Hessian}
\end{align}
the Hessian for observation $t$.

ABT showed that under regularity conditions, the QMLE is asymptotically normally distributed, 
$\sqrt{T}\left(\boldsymbol{\widehat{\theta}}_T-\boldsymbol{\theta}_0\right) \overset{D}{\to} N\left(\mathbf{0},\mathbf{B}^{-1}\mathbf{A}\mathbf{B}^{-1}\right), \text{ as } T\to\infty$. The expressions for
$\mathbf{A}$ and $\mathbf{B}$ are integrals of the stationary approximation of the score and Hessian (see ABT, Theorem 2).  The derivations of the asymptotic properties in ABT rely on the general theory for locally stationary processes developed by \textcite{dahlhaus2019towards}.

\subsection{Lagrange multiplier test}
Unidentified parameters under the null hypothesis form a complication in parametric tests of smooth structural change. A solution to the identification problem was developed by \textcite{LST88b}, who use a Taylor expansion around the null hypothesis to approximate the unidentified nonlinearity. In connection with GARCH models, many authors have used this method. See, for example, \textcite{MedeirosW09} who test for a flexible coefficient GARCH model and \textcite{AmadoT15} who develop misspecification tests for MTV-GARCH.

Let $\boldsymbol{\theta}=(\boldsymbol{\theta}_{1}^{\intercal},\boldsymbol{\theta}_{2}^{\intercal},)^{\intercal}$, where $$\boldsymbol{\theta}_1=(\alpha_{01},\ldots\alpha_{0L}, \gamma_1,\ldots,\gamma_L, c_{1},c_{2},\ldots,c_{L})^{\intercal},$$ and $$\boldsymbol{\theta}_2=(\alpha_0, \alpha_1,\ldots, \alpha_p,\beta_1,\ldots,\beta_q)^{\intercal},$$ be a vector containing the parameters of the model (\ref{eq:GARCH})--(\ref{eq:lfunct}).

Additive misspecification is defined as omitting a nonlinear function of rescaled time that enters the true model additively. Let $$\boldsymbol{\theta}_3=(\alpha_{0,L+1}, \gamma_{L+1}, c_{L+1})^{\intercal}.$$ Consider the following specification: $        \sigma^2_t(\boldsymbol{\theta}_1, \boldsymbol{\theta}_2, \boldsymbol{\theta}_3) = h_t(\boldsymbol{\theta}_1, \boldsymbol{\theta}_2, \boldsymbol{\theta}_3) + g_t(\boldsymbol{\theta}_1) + g_t(\boldsymbol{\theta}_3)$, where
$$
h_t(\boldsymbol{\theta}_1, \boldsymbol{\theta}_2, \boldsymbol{\theta}_3)=\alpha_{0} + \sum_{j=1}^{p}\alpha_{j}X_{t-j}^{2}+\sum_{i=1}^{q}\beta_{i}\sigma_{t-i}^{2},
$$ 
$$g_t(\boldsymbol{\theta}_1) = \sum_{l=1}^{L}\alpha_{0l}G\left(\frac{t}{T},\gamma_l,c_l\right) \quad \text{and} \quad
g_t(\boldsymbol{\theta}_3) = \alpha_{0(L+1)}\left\{ G\left(\frac{t}{T},\gamma_{L+1},c_{L+1}\right)-\frac{1}{2}\right\}.$$
Testing for additive misspecification implies the null hypothesis $H_0$: $ \alpha_{0(L+1)}=0,$ or equivalently, $H_0$: $\gamma_{L+1}=0$.
In the former case, $\gamma_{L+1}$ and $c_{L+1}$ and in the latter, $\alpha_{0(L+1)}$ and $c_{L+1}$ are unidentified nuisance parameters. To overcome this problem, we consider the latter null hypothesis and approximate the unidentified nonlinearity by a third-order Taylor series expansion around $H_0$:
\begin{equation}
g_t(\boldsymbol{\theta}_3) = \boldsymbol{\delta}^{\intercal}
\boldsymbol{\tau}_{t} + R_{t}, \label{eq:unidLM2}
\end{equation}
where $\boldsymbol{\delta} = (\delta_{1},\delta_{2},\delta_{3})^{\intercal}$, $\boldsymbol{\tau}_{t} = (t/T, (t/T)^2, (t/T)^3)^{\intercal}$ and $R_{t}$ is the remainder to be ignored. Since $\boldsymbol{\delta} = \boldsymbol{0}$ if and only if $\gamma_{L+1} = 0$, the new null hypothesis is  $H_{0}'$: $\boldsymbol{\delta} = \boldsymbol{0}$, tested against $H_{1}'$: $\boldsymbol{\delta} \neq \boldsymbol{0}$. By this approximation, the testing problem is transformed into testing a linear hypothesis, so standard asymptotic inference applies. 

To construct an LM test, we require the derivatives of $\sigma_{t}^{2}$ with respect to $\boldsymbol{\theta}_1$, $\boldsymbol{\theta}_2$ and $\boldsymbol{\delta}$. For example, for the GARCH$(1,1)$ model, the approximation (\ref{eq:unidLM2}) yields \begin{align}
 \frac{1}{\sigma^2_t}\frac{\partial\sigma^2_t}{\partial \boldsymbol{\delta}}
 &=\frac{1}{\sigma^2_t}\left(\boldsymbol{\tau}_{t} + \beta_1\boldsymbol{\tau}_{t-1} + \beta_1^2\boldsymbol{\tau}_{t-2} + \cdots + \beta_1^j\boldsymbol{\tau}_{t-j} + \cdots \right). \label{eq:dft}
\end{align}
In the implementation of the test, the derivatives are truncated at $t-j=0$ (see Appendix A), meaning that the sum in (\ref{eq:dft}) ranges at most from $j=0$ to $j=t.$ The test thus takes the recursive nature of the GARCH process into account; see \textcite{halungaorme2009}.

We now proceed to construct the LM test. In terms of the previous notation, the score for observation $t$ becomes $
\boldsymbol{s}_{t}(\boldsymbol{\widehat{\theta}}) := \boldsymbol{s}_{t}(\widehat{\boldsymbol{\theta}}_{1T},\widehat{\boldsymbol{\theta}}_{2T},\boldsymbol{0})=(1/2)(\widehat{
Z}_{t}^{2}-1)\left(\widehat{\boldsymbol{s}}_{1t}^{\intercal},  
\widehat{\boldsymbol{s}}_{2t}^{\intercal}, \widehat{\boldsymbol{s}}_{3t}^{\intercal}\right)^{\intercal}$,
where $\widehat{\boldsymbol{s}}_{nt}=(1/\widehat{\sigma}_{t}^{2})\left. (\partial
\sigma_{t}^{2}/\partial \boldsymbol{\theta} _{n})\right\vert _{\boldsymbol{\theta} = \boldsymbol{\widehat{\theta}}}$,
for $n = 1,2,3.$ The derivative $\widehat{\boldsymbol{s}}_{3t}$ is approximated as in (\ref{eq:dft}). Let $\mathbf{s}_{T}(\boldsymbol{\widehat{\theta}})=(1/T)\sum_{t=1}^{T} \mathbf{s}_t(\boldsymbol{\widehat{\theta}})$. Assuming $Z_t\sim \mbox{NID}(0,1)$, the expected Hessian can be consistently estimated under $H_{0}$ by $\boldsymbol{H}_T(\widehat{\boldsymbol{\theta}}) = (1/T)\sum_{t=1}^T\boldsymbol{s}_{t}(\boldsymbol{\widehat{\theta}}) \boldsymbol{s}^{\intercal}_{t}(\boldsymbol{\widehat{\theta}})$. By consistency and asymptotic normality of the estimator $\widehat{\boldsymbol{\theta}}_T$, we obtain for the LM statistic under $H_{0}'$:
\begin{equation*}
    LM = \mathbf{s}_{T}^{\intercal}(\boldsymbol{\widehat{\theta}})\mathbf{H}_{T}(\boldsymbol{\widehat{\theta}})^{-1} \mathbf{s}_{T}(\boldsymbol{\widehat{\theta}}) \overset{D}{\to} \chi^{2} (3), \text{ as } T\to\infty.
\end{equation*}

The test statistic can be made robust against violations of distributional assumptions. Appendix A details how to calculate the LM statistic and a robust version of it by auxiliary regressions.

We end this section by some remarks on the test and the testing procedure. As already pointed out in Section 2.1, the ATV-GARCH model and MTV-GARCH model of \textcite{AmadoT13} are asymptotically equivalent. As far as testing is concerned, a computational advantage of the ATV-GARCH model is that its parameters can be estimated by QML in a single step, whereas the parameters of the MTV-GARCH model have to be estimated by an iterative procedure called estimation by parts, detailed in \textcite{SongFK05}. When the null is GARCH, the differences between the tests are small because no estimation of a transition function is involved. In testing $L$ transitions against $L+1$, our tests are numerically simpler and faster to compute than the ones in Amado and Teräsvirta (2017), since the numerically more demanding estimation by parts is avoided.

Since our test of GARCH against ATV-GARCH is designed to detect smooth change in one of the GARCH parameters, it is similar to the test of parameter constancy in \textcite{SLTT02}. The LM test in \textcite{AmadoT15} and our LM test use the same third-order Taylor approximation (\ref{eq:dft}). The columns of the score corresponding to $\boldsymbol{\delta}$ in the additive and multiplicative tests are linearly dependent up to the time-varying factor $(1/\sigma^2_t)$, which is not present in the test in \textcite{AmadoT15}. Our LM test has therefore power against both additive and multiplicative misspecification. As already mentioned, our LM tests take the recursion in the conditional heteroskedasticity under the alternative into account whereas the tests in \textcite{SLTT02} and \textcite{AmadoT15} do not.

The test implies a sequential specific to general procedure as in e.g. \textcite{MedeirosW09} and Amado and Ter{\"a}svirta (2017). Start with a linear model (zero transition functions) and tests against an alternative hypothesis of one transition function. If the test rejects, fit a model with one transition and continue by testing it against the model with two transitions. Proceed until the first non-rejection. This way, the tests are used to determine the number of transition functions (regimes) and may thus be viewed as both specification and misspecification tests.

\section{Simulation study}
In this section, we investigate the finite-sample properties of the LM test and its robust version against an additive time-varying GARCH model by simulation. 
\subsection{Simulation design}
 We have programmed routines in \texttt{R} to simulate data, fit the models under the null hypothesis and compute the test statistics. ML estimation is done numerically using the solver \texttt{Solnp} by \textcite{Ye1987}, implemented in the \texttt{R} package \texttt{Rsolnp} by \textcite{Rsolnp}. See Appendix B for details. We use time series lengths $T = 1000$, $2500$, $5000$ and a burn-in period of $200$ observations. The number of Monte Carlo replications is $5000$.

 \begin{table}[h] 
\begin{center}
\caption{DGPs in the simulation study.}
\resizebox{0.7\linewidth}{!}{
\begin{tabular}{l c c c c c c c c c c c c}

\toprule

& &  \multicolumn{3}{c}{GARCH} &   &\multicolumn{6}{c}{Logistic transition function} \\
\cmidrule{3-5} \cmidrule{7-12}
&  & $\alpha_{0}$ & $\alpha_1$ & $\beta_1$ &  & $\alpha_{01}$ & $\gamma_1$ & $c_1$ & $\alpha_{02}$ & $\gamma_2$ & $c_2$ & Shape\\

\hline
DGP 1 &     & $0.1$  & $0.1$   & $0.85$   & & $-$ & $-$ & $-$   & $-$   & $-$  & $-$  & $-$\\
     \rowcolor{gray!20}
DGP 2  &    & $0.05$  & $0.05$   & $0.9$   &  & $-$ & $-$ & $-$   & $-$   & $-$  & $-$ & $-$ \\
DGP 3 &    & $0.005$  & $0.05$   & $0.8$   &  & $-$ & $-$  & $-$   & $-$   & $-$  & $-$  & $-$\\
     \rowcolor{gray!20}
DGP 4   &   & $0.005$  & $0.05$   & $0.8$   &  & $0.015$ & $10$  & $0.5$   & $-$   & $-$  & $-$ & \begin{sparkline}{10}
\spark 0.10 0.02 0.20 0.05 0.30 0.12 0.40 0.27 0.50 0.50 0.60 0.73 0.70 0.88 0.80 0.95 0.90 0.98 1.00 0.99 /
\end{sparkline} \\
DGP 5  &    & $0.005$  & $0.05$   & $0.8$   &  & $0.005$& $10$  & $0.5$   & $-$   & $-$  & $-$ &  \begin{sparkline}{10}
\spark 0.10 0.02 0.20 0.05 0.30 0.12 0.40 0.27 0.50 0.50 0.60 0.73 0.70 0.88 0.80 0.95 0.90 0.98 1.00 0.99 /
\end{sparkline}\\
     \rowcolor{gray!20}
DGP 6 &    & $0.005$  & $0.05$   & $0.8$   &  & $0.0025$ & $10$ & $0.5$   & $-$   & $-$  & $-$ & \begin{sparkline}{10}
\spark 0.10 0.02 0.20 0.05 0.30 0.12 0.40 0.27 0.50 0.50 0.60 0.73 0.70 0.88 0.80 0.95 0.90 0.98 1.00 0.99 /
\end{sparkline} \\
DGP 7 &    & $0.005$  & $0.05$   & $0.8$  &  & $0.015$ & $5$  & $0.5$   & $-$   & $-$  & $-$ & \begin{sparkline}{10}
\spark 0.10 0.12 0.20 0.18 0.30 0.27 0.40 0.38 0.50 0.50 0.60 0.62 0.70 0.73 0.80 0.82 0.90 0.88 1.00 0.92 /
\end{sparkline}  \\
     \rowcolor{gray!20}
DGP 8   &    & $0.005$  & $0.05$   & $0.8$  &  & $0.005$ & $5$  & $0.5$   & $-$   & $-$  & $-$ & \begin{sparkline}{10}
\spark 0.10 0.12 0.20 0.18 0.30 0.27 0.40 0.38 0.50 0.50 0.60 0.62 0.70 0.73 0.80 0.82 0.90 0.88 1.00 0.92 /
\end{sparkline}   \\
DGP 9  &   & $0.005$  & $0.05$   & $0.8$   &  & $0.0025$& $5$  & $0.5$   & $-$   & $-$  & $-$ & \begin{sparkline}{10}
\spark 0.10 0.12 0.20 0.18 0.30 0.27 0.40 0.38 0.50 0.50 0.60 0.62 0.70 0.73 0.80 0.82 0.90 0.88 1.00 0.92 /
\end{sparkline}  \\
     \rowcolor{gray!20}
DGP 10  &    & $0.005$  & $0.05$   & $0.8$   & & $0.01$ & $10$ & $0.25$   & $-0.01$   & $10$  & $0.75$ & \begin{sparkline}{10}
\spark 0.10 0.18 0.20 0.37 0.30 0.61 0.40 0.79 0.50 0.85 0.60 0.79 0.70 0.61 0.80 0.37 0.90 0.18 1.00 0.08 /
\end{sparkline} \\
DGP 11  &     & $0.005$  & $0.05$   & $0.8$  &  & $0.005$ & $10$  & $0.25$   & $0.005$   & $10$  & $0.75$  & \begin{sparkline}{10}
\spark  0.10 0.02 0.20 0.13 0.30 0.37 0.40 0.48 0.50 0.50 0.60 0.52 0.70 0.63 0.80 0.87 0.90
0.98 1.00 1.00 /
\end{sparkline}  \\

\hline
\multicolumn{13}{l}{\scriptsize{\textbf{Note:} The "Shape" column contains rough depictions of the transition function.}}
\end{tabular}}
\label{tab:dgps}
\end{center}
\end{table}

In the size simulations we investigate the size of the tests of a standard GARCH model against an ATV-GARCH model with one transition and an ATV-GARCH model with one transition against two. In the power simulations we consider the following cases: testing GARCH against ATV-GARCH with one transition, testing GARCH against ATV-GARCH with two transitions and testing ATV-GARCH with one transition against ATV-GARCH with two transitions.  

The data-generating process (DGP) is given by $X_t = \sigma_t Z_t$,
where $Z_t\sim \text{NID}(0,1)$, unless stated otherwise. DGPs 1--3 are designed to investigate the size of the test of a standard GARCH model against an ATV-GARCH model with one transition. The DGPs differ in terms of the magnitude of the ARCH coefficient $\alpha_1$ and the persistence measured by $\alpha_1 + \beta_1$. The parameter values are contained in Table \ref{tab:dgps}. The DGPs generate observations from distributions which are mildly leptokurtic. The values of excess kurtosis for DGPs 1--3 are 0.774, 0.162 and 0.055, respectively. In addition we consider a DGP, called DGP 3t, where the parameters are as in DGP 3, but the errors are $t$-distributed with $5$ degrees of freedom. To simulate the test of one transition against two, we employ DGPs 4--9, where $\sigma^2_{t}= 0.005 + \alpha_{01}G(t/T; \gamma_{1}, c_{1}) +  0.05X^2_{t-1}+0.80\sigma^2_{t-1}$. In these DGPs the GARCH specification is as in DGP 3. We choose a low level of persistence ($\alpha_1 + \beta_1 = 0.85$) in order to focus on how the size of the test of one transition against two transitions depends on $\alpha_{01}$, $\gamma_{1}$ and $c_{1}$. In DGPs 4 and 7, the intercept increases fourfold from $0.005$ for $G=0$ to $0.020$ for $G=1$. This increase is twofold in DGPs 5 and 8, and $1.5$-fold in DGPs 6 and 9. The values for the slope parameter are $\gamma_{1}=5$ and $10$, and the value for $c_{1}$ is $0.5$. For the alternative with two transitions, we introduce two additional DGPs. DGP 10 contains two transition functions with a sign change and DGP 11 two transitions of the same sign. The first transition is centred at $c_{1}=0.25$ and the second one at $c_{2}=0.75$.

\subsection{Size results}

The results from the size simulations are reported using size discrepancy plots (\cite{davidson1998graphical}), which show the size discrepancy defined as the empirical size minus the nominal size plotted against the nominal size from $0.1\%$ to $10\%$. The simulated rejection probabilities at the $1\%$, $5\%$ and $10\%$ significance levels are reported in Table \ref{tab:size} in Appendix B.

\begin{figure}[h]
    \centering
    \caption{Simulated size of the LM test. The null model is GARCH.}
    \label{fig:dgp1233t}
    \includegraphics[width=0.82\linewidth]{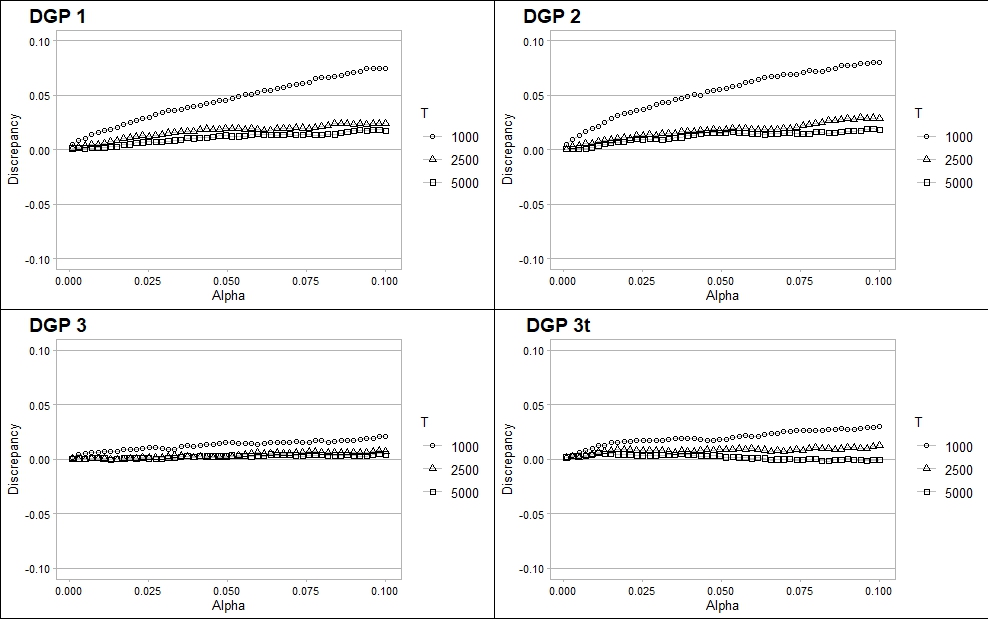}
    \caption*{\footnotesize{Note: Circle: $T=1000$, Triangle: $T=2500$, Square: $T=5000$.}}
\end{figure}

\begin{figure}[h]
    \centering
    \caption{Simulated size of the robust LM test. The null model is GARCH.}
    \label{fig:rdgp1233t}
    \includegraphics[width=0.82\linewidth]{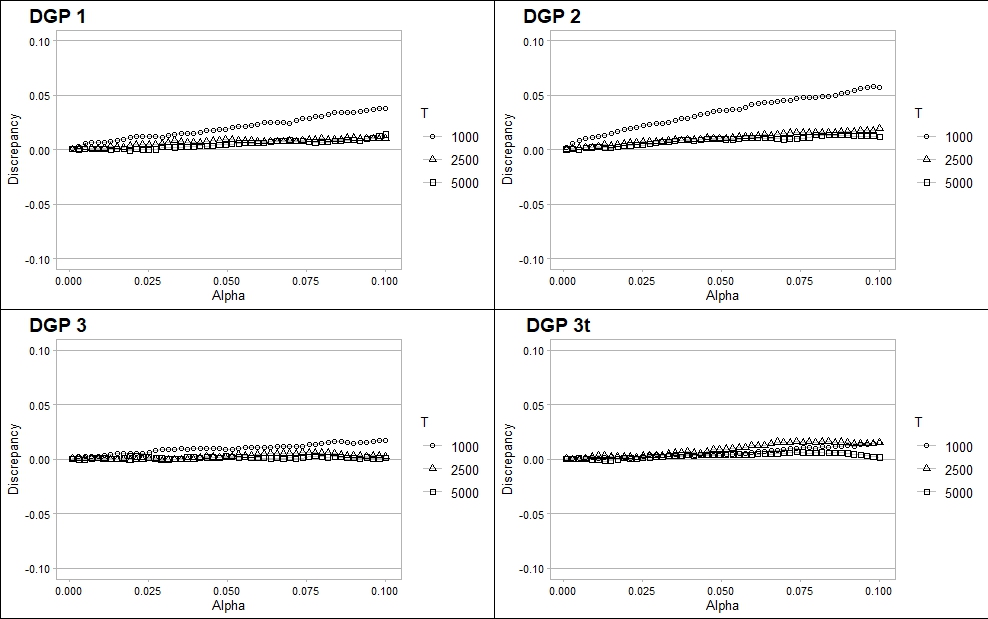}
    \caption*{\footnotesize{Note: Circle: $T=1000$, Triangle: $T=2500$, Square: $T=5000$.}}
\end{figure}

The LM test of a standard GARCH model against an ATV-GARCH model with one transition exhibits positive size discrepancies in DGPs 1--3t; see Figure \ref{fig:dgp1233t}. We find that when the persistence $\alpha_1 + \beta_1$ increases, the size discrepancy becomes larger and the empirical size of the test converges slower to its nominal level. This is what we would expect, since when persistence is high, volatility clusters tend to become longer and the test may misinterpret such clusters as smooth structural change. This is true in particular when there is a large cluster towards the end of the time series. The effect dissipates in longer time series as the emergence and disappearance of volatility clusters become more frequent. Consequently, the test tends to be slightly oversized for small and moderate sample sizes when persistence is high. The results for the robust LM test are shown in Figure \ref{fig:rdgp1233t}. It has better size properties than the LM test, in particular for small and moderate values of $T$. We find that the robust LM test is well-sized except for the smallest sample size $T=1000$ and when persistence is high.
Figures \ref{fig:dgp1233t} and \ref{fig:rdgp1233t} confirm that for $T=5000$, the size of both tests is close to the nominal level. We conclude that the asymptotic distribution is a good approximation to the finite-sample distribution for $T \geq 2500$. The LM test is slightly oversized in DGP 3t for the smaller sample sizes $T=1000$ and $2500$ when the error distribution is erroneously assumed to be standard normal. The test is well-sized for $T=5000$ even when the error distribution is misspecified.

\begin{figure}[h]
    \centering
    \caption{Simulated size of the robust LM test. The null model is ATV-GARCH.}
    \label{fig:rdgp456789}
    \includegraphics[width=0.82\linewidth]{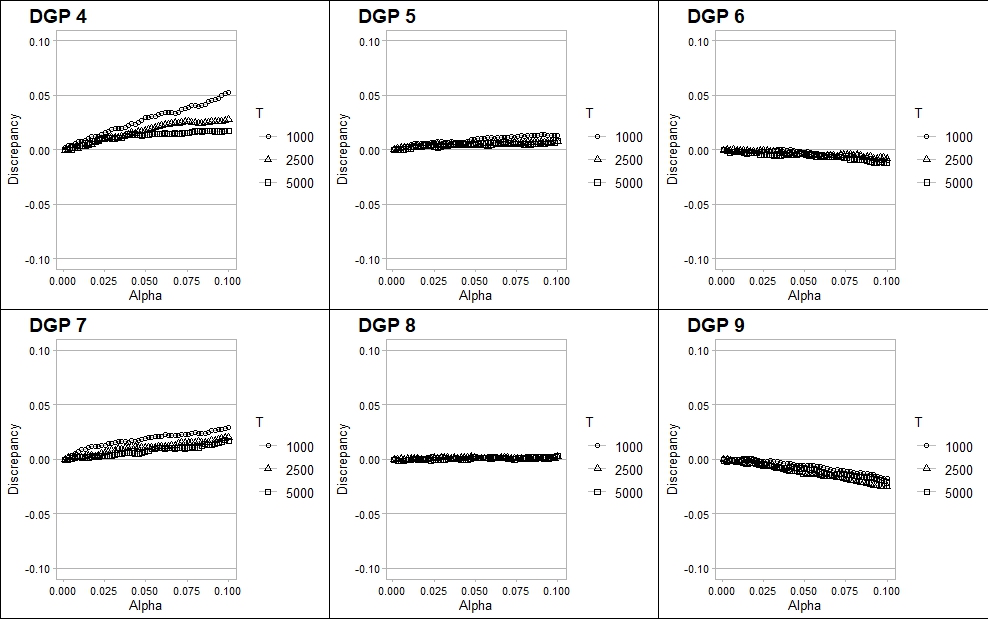}
    \caption*{\footnotesize{Note: Circle: $T=1000$, Triangle: $T=2500$, Square: $T=5000$.}}
\end{figure}

The size results for the robust LM test of testing an ATV-GARCH model with one transition against two in DGPs 4--9 are presented in Figure  \ref{fig:rdgp456789}. The test is oversized when the transition is large. On the other hand, the test is slightly undersized when the transition is small. Size distortion decreases with $T$.  A similar pattern is found for the non-robust LM test. To save space, we only show the results for the robust version of the test.

In conclusion, both the LM test and the robust version are well-sized both when the alternative is a standard GARCH model and an ATV-GARCH model with an estimated transition function. Since the empirical size of the robust LM test is never worse than that of the LM test, it is advisable to use the former. 

\subsection{Power simulations}

Power is reported in the form of size-power curves, which show 
power against the nominal size from $0.1\%$ to $10\%$. To save space, we only show the results for the robust LM test. 
The power curves are not size-adjusted because the size of the robust LM test is close to the nominal level. The simulated powers at the $5\%$ and $10\%$ significance levels are reported in Table \ref{tab:power} in Appendix B.

\begin{figure}[h]
    \centering
    \caption{Simulated power functions of the robust LM test. The true model is ATV-GARCH with one transion.}
    \label{fig:rpower456789}
    \includegraphics[width=0.82\linewidth]{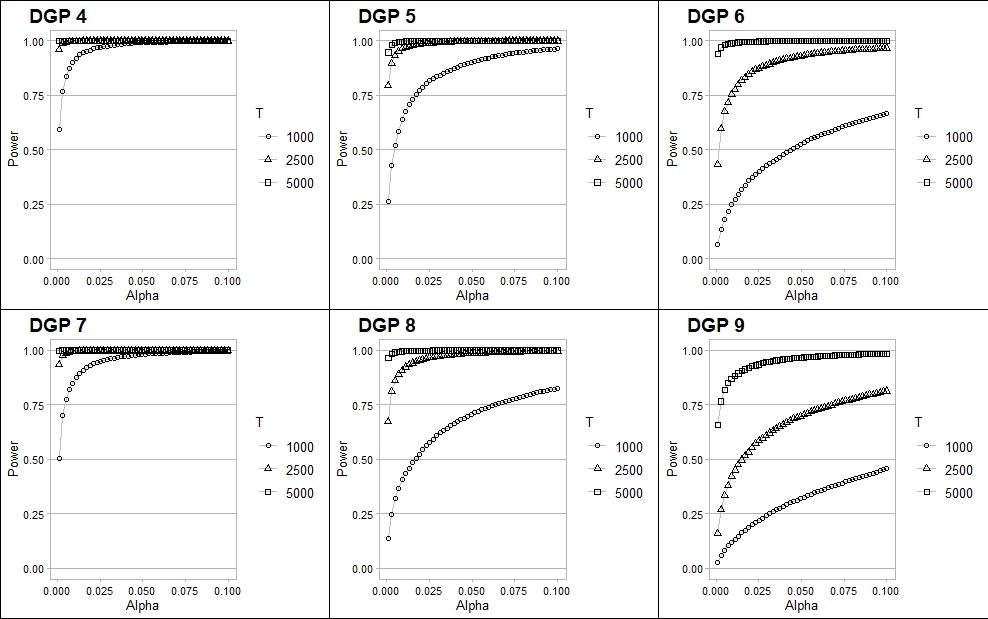}
    \caption*{\footnotesize{Note: Circle: $T=1000$, Triangle: $T=2500$, Square: $T=5000$.}}
\end{figure}

We begin with the test of GARCH against ATV-GARCH with one transition. The size-power curves of the robust LM test can be found in Figure \ref{fig:rpower456789}. As expected, the power is increasing in $\alpha_{01}$ and $\gamma$. The power is close to $1$ at the nominal level $5\%$ in DGPs 4 and 7 (the intercept increases by a factor of four) and DGPs 5 and 8 (the intercept increases by a factor of two), except for the smallest sample size $T=1000$. The only case where the power is below $50\%$ is in DGP 9 (the intercept increases by a factor of $1.5$ and $\gamma_1=5$) and $T=1000$. For a small transition ($\alpha_{01}$ small) a longer time series is needed to detect it. Taking DGP 9 with $\alpha_{01}=0.0025$ as an example, the power of the robust LM test is $32\%$ when $T=1000$, $70\%$ when $T=2500$ and $97\%$ when $T=5000$. The LM test (not reported) and its robust version are about equally powerful and the differences in empirical power in favour of the LM test are due to the former being slightly oversized.

\begin{figure}[h]
    \centering
    \caption{Simulated power functions of the robust LM test. The true model is ATV-GARCH with two transions.}
    \label{fig:rdgp1011}
    \includegraphics[width=0.82\linewidth]{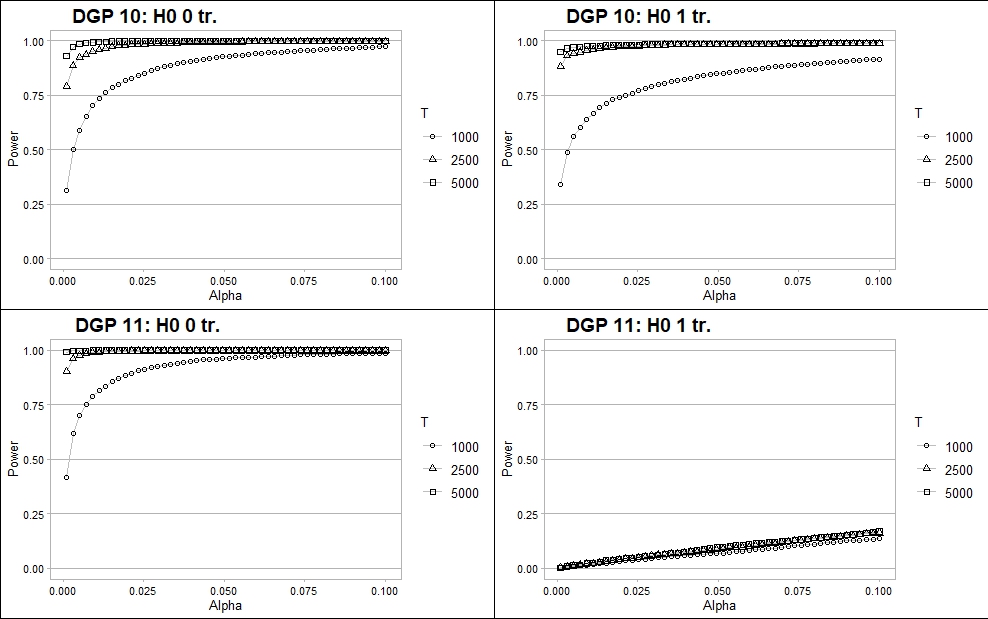}
    \caption*{\footnotesize{Note: Circle: $T=1000$, Triangle: $T=2500$, Square: $T=5000$.}}
\end{figure}

DGPs 10 and 11 contain two transitions with either a sign change (DGP 10) or two transitions of the same sign (DGP 11). We would expect power to be high in DGP 10 both against one and two transitions but lower when testing one transition against two in DGP 11. The size-power curves can be found in Figure \ref{fig:rdgp1011}. Both the test against $L=1$ and the one against $L=2$ have high power when there is a sign change in the transition function. Power is greatly reduced when there are two transitions of the same sign. In this case, fitting one transition to a weakly monotonic curve containing two transitions removes enough nonlinearity to diminish the power of the second test.

Summing up, the power simulations show that the tests have high power to detect smooth transition in the unconditional volatility already when the sample size is moderate. 

\section{Empirical application}

In this section we model the volatility of the daily log-returns on the VIX index for the entirety of its (back-calculated) existence by an ATV-GARCH model.

\begin{figure}[h]
    \caption{Daily log-returns on the VIX index from 3 January 1990 to 7 April 2022}
    \label{fig:vix}
    \begin{subfigure}{0.5\textwidth}
    \centering
        \includegraphics[width=0.9\linewidth]{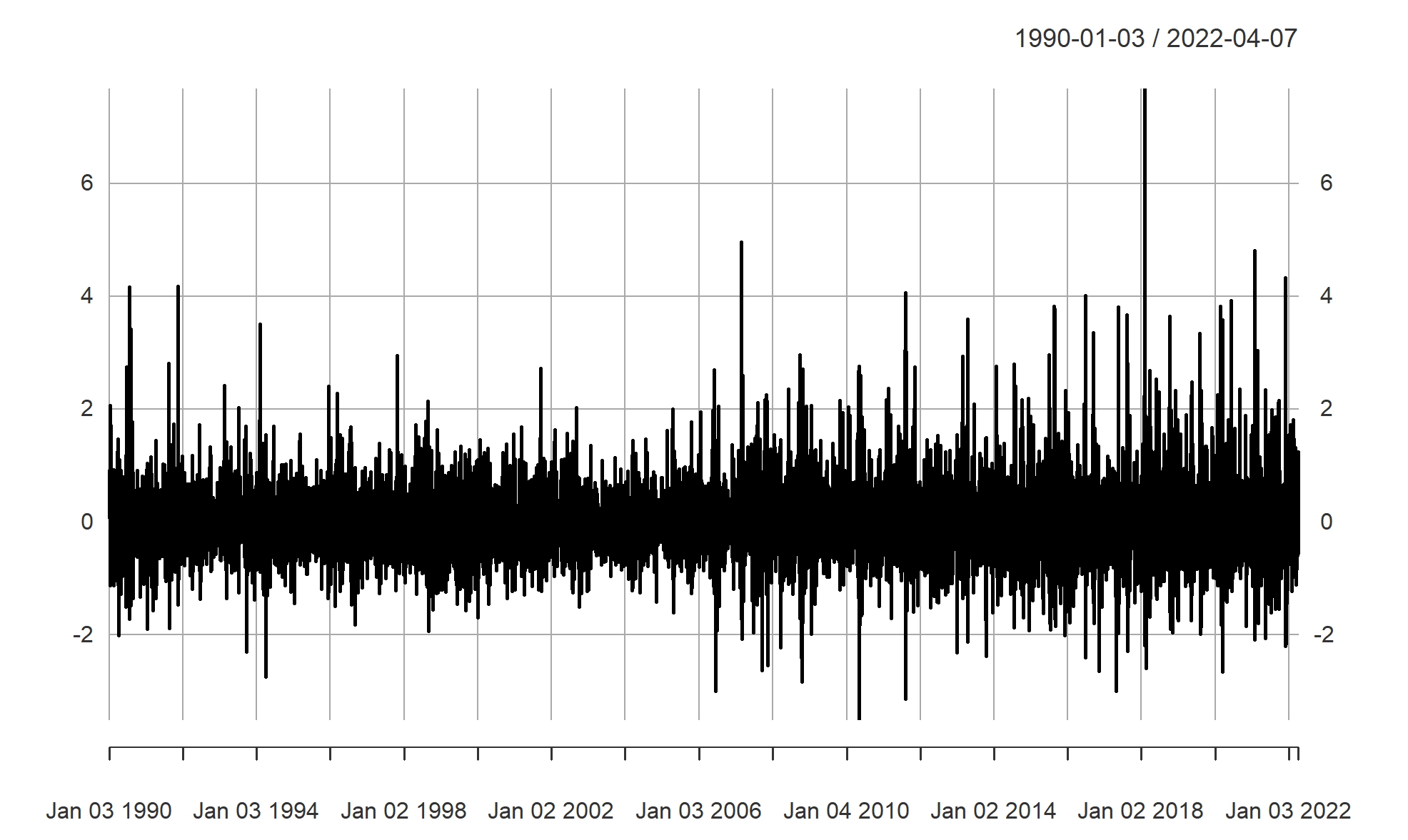}
        \caption{Daily log-returns}
 \label{fig:vixr}
    \end{subfigure}%
    ~
    \begin{subfigure}{0.5\textwidth}

        \includegraphics[width=0.9\linewidth]{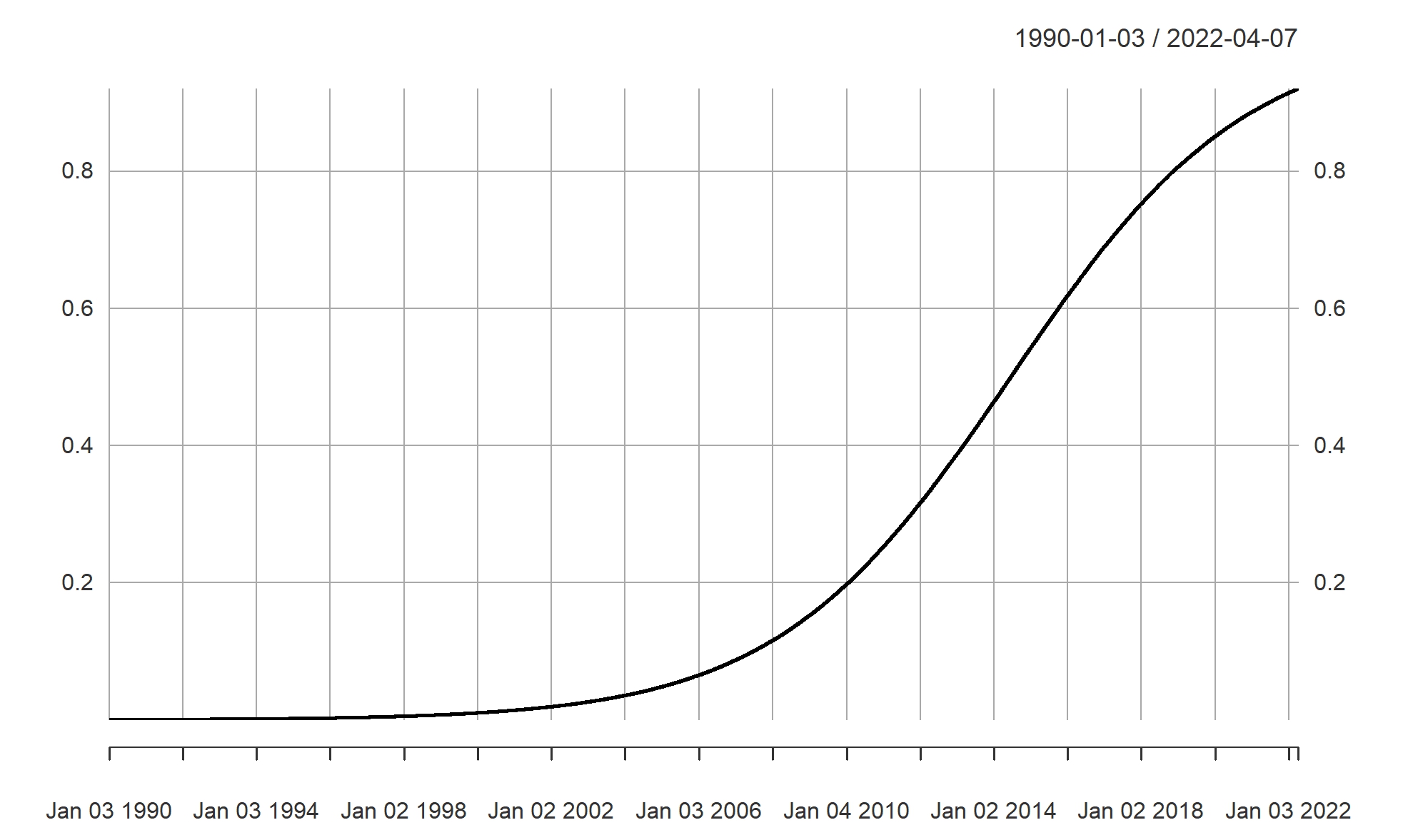}
        \caption{The estimated logistic transition function}
 \label{fig:vixgt}
    \end{subfigure}%
   \end{figure}

The VIX index is an index of volatility implied by option prices. Figure \ref{fig:vixr} presents its daily log-returns for the entirety of its (back-calculated) existence from 3 January 1990 to 7 April 2022. Data are from the Chicago Board Option Exchange\footnote{Cboe, downloaded on 9 April, 2025.}. The series contains $8127$ observations. For numerical reasons, the returns are multiplied by $10$. Summary statistics for the daily log-returns can be found in the first panel of Table \ref{tab:empapp}. The robust measures of skewness and kurtosis are the ones promoted by \textcite{kimwhite2004}. Figure \ref{fig:vixr} suggests that the amplitude of the clusters may begin to increase around 2007, which, if true, would coincide with the 2007--2008 financial crisis. This conjecture will be investigated by the ATV-GARCH model.

\subsection{Testing for additive misspecification}

We begin by testing GARCH$(1,1)$ against ATV-GARCH$(1,1)$. Since the series contains more than $8000$ observations, we regard a $p$-value below $0.001$ as convincing evidence against the null hypothesis of constant conditional variance (standard GARCH). The values of the LM statistic ($52.1$) and the robust LM statistic ($29.8$) lie above this critical level, see Table \ref{tab:empapp}. Interestingly, the persistence of the standard GARCH is quite low: $\widehat{\alpha}_1 + \widehat{\beta}_1 = 0.891$. This is different from the situations usually encountered in practice. Typically, the persistence of the null model is close to unity (the border where the variance becomes infinite), and the smooth deterministic component lowers it considerably. See, for example, \textcite{Feng04} or \textcite{AmadoT15}.

The persistence in the estimated ATV-GARCH$(1,1)$ model is indeed lower than that from the GARCH$(1,1)$, but as seen from
Table \ref{tab:empapp} ($\widehat{\alpha}_1 + \widehat{\beta}_1 = 0.858$), the decrease  is not very large. Nevertheless, it is remarkable that the test does clearly detect the amplitude change, so a nonlinear ATV-GARCH model can be sucessfully fitted to the data. From Table \ref{tab:empapp} it is further seen that the test of one transition against two does not favour an additional transition, so the ATV-GARCH model with one transition remains the final one. The full estimation results can be found in the last panel of Table \ref{tab:empapp}. The $t$-statistics of the coefficient estimates are based on standard errors computed using the inverse of the numerically computed Hessian. These are valid under the assumption of normally distributed errors. We also report robust $t$-statistics, which are calculated using the sandwich form of the variance-covariance matrix with numerically computed variance-covariance matrix of the score.
\begin{table}[h]
\centering
\caption{Summary statistics, LM tests and parameter estimates for the VIX returns.}
\resizebox{0.9\columnwidth}{!}{
\begin{tabular}{lccccccccc} 
\toprule
\textbf{Statistics}   & ~ & ~ & ~ & ~ & ~ & ~ & ~ \\ 
Series & Mean & Sd & Med & Min & Max & Skew & Kurt & R Skew  & R Kurt\\ 
\hline
VIX & $0.000$ & $0.674$ & $-0.038$ & $-3.506$ & $7.682$ & $0.962$ & $9.403$ &$0.03$ & $0.168$ \\ 
\bottomrule
\textbf{Testing} & & & & & & & & &\\ 
Null & $LM$ & $p$-value & $LMr$ & $p$-value & $\widehat{\alpha}_1$ &  $\widehat{\beta}_1$ & $\widehat{\alpha}_1 + \widehat{\beta}_1$ & & \\ \hline
$0$ transitions & $52.080$ & $0.000$ & $29.779$ &   $0.000$ & $0.131$ & $0.760$ &  $0.891$ \\
$1$ transition & $4.868$ & $0.182$ &  $4.287$ & $0.232$ & $0.126$ &  $0.732$ & $0.858$ & &\\        
\bottomrule
\textbf{Final model} 
& & & & & & & & &\\ 
Coeff & Estim & se & R se & Coeff & Estim & se & R se & &\\ 
$\widehat{\alpha}_{0}$ & $0.047$ & $0.005$ &   $0.010$ & $\widehat{\eta}_{1}$ & $0.910$ & $0.023$ & $0.043$ & & \\   
$\widehat{\alpha}_{1}$ & $0.126$ & $0.011$ &   $0.017$ & $\widehat{c}_1$ & $0.758$ & $0.070$ & $0.165$ & &\\ 
$\widehat{\beta}_{1}$ & $0.732$ & $0.022$ & $0.039$ & $\widehat{\alpha}_{01}$ & $0.069$ & $0.018$ & $0.043$ & &\\
\bottomrule

\end{tabular}
}
\raggedright{{\\ \scriptsize{\ \textbf{Note:} \\ \ We have estimated $\gamma_1$ using the parameter transformation $\gamma_1 = \eta_1 / (1-\eta_1)$. The reverse transformation yields $\widehat{\gamma}_1 \approx 10.15$. \\ \ The returns are multiplied by 10. All values are rounded to three decimals. \\ \ Mean is the average value of the series over the sample period, SD is the standard deviation and Med is the \\ \ median. Min and Max are the minimum and maximum values, respectively. Skew is the skewness and Kurt is the \\ \ kurtosis. "R" signifies a robust version of the statistic. Robust skewness is Bowley's skewness and robust kurtosis is (centered) Moor's kurtosis. \\ \ }}} 
\label{tab:empapp}
\end{table}

The estimate of the shape parameter $\gamma_1$ is $10.15$, which 
corresponds to a slow transition taking place over a period of several years. This results in a gradual, but noticeable, level-change in the conditional variance centred at about three-quarters of the length of the time series ($\widehat{c}_1=0.758$). The estimated transition function is depicted in Figure \ref{fig:vixgt}. It is smooth and begins its increase around 2007. The coefficient $\widehat{\alpha}_{01}=0.069$ can be compared with the intercept $\widehat{\alpha}_{0} = 0.047$. Although the increase is slow, it is large in magnitude. The daily unconditional volatility increases from about $5.8\%$ towards $9.0\%$, a $1.6$-fold increase.

\section{Conclusion}

In this paper, the focus is on specifying the time-varying intercept in the ATV-GARCH model introduced by ABT. In order to avoid estimation of unidentified models, the form of the intercept is specified by a sequence of tests, moving from specific to general. The first test is the test of the standard stationary GARCH model against an ATV-GARCH model with one transition. If the null hypothesis is rejected, the alternative becomes the next null model, and is tested against an ATV-GARCH with two transitions. Testing proceeds until the first non-rejection. To circumvent the identification problem, all tests are based on approximating the alternative. A standard Lagrange multiplier test and a robust version of it are derived. Both test statistics have the same asymptotic $\chi^{2}$-distribution under the null hypothesis. Simulations show that both tests are well-sized for moderate and large samples. The application to the VIX index provides an interesting result. Judging from the estimated persistence (the sum of the GARCH parameters is $\widehat{\alpha}_1 + \widehat{\beta}_1 \approx 0.89$), stationary GARCH$(1,1)$ appears to be an adequate choice. Nevertheless, the test rejects this model in favour of ATV-GARCH. The estimated ATV-GARCH$(1,1)$ model indicates that the unconditional variance begins a slow ascent around the 2007--2008 financial crisis and continues to increase towards the end of the observation period.
\clearpage
\printbibliography
\newpage
\section*{Appendix}

Appendix A contains computational details of the LM tests. Appendix B contains the simulated rejection probabilities and powers of the LM tests at the $5\%$ and $10\%$ significance levels to examine the quality of the asymptotic approximation of the distribution of the LM statistics. 

\subsection*{A Computational details}

\subsection*{Calculation of derivatives}
The partial derivatives in (\ref{eq:score}) have a recursive structure. For the standard GARCH case, this structure is considered in detail by \textcite{FiorentiniCP96}. For example, for an ATV-GARCH$(1,1)$ model with one transition function we have \begin{align*}
\frac{\partial \sigma_{t}^{2}}{\partial \boldsymbol{\theta}_{1}}=& \frac{\partial}{\partial \boldsymbol{\theta}_{1}}\left\{\alpha_{0}+\alpha_{01} G_{1}\left(t / T ; \gamma_{1}, \mathbf{c}_{1}\right)+\alpha_{1} X_{t-1}^{2}+\beta_{1} \sigma_{t-1}^{2}\right\} \\
=&\left(G_{1}\left(t / T ; \gamma_{1}, \mathbf{c}_{1}\right),\right.\\
& \alpha_{01} G_{1}^{2}\left(t / T ; \gamma_{1}, \mathbf{c}_{1}\right) \exp \left(-\gamma_{1}\left(t / T-c_{11}\right)\right)\left(t / T-c_{11}\right), \\
&\left.\alpha_{01} G_{1}^{2}\left(t / T ; \gamma_{1}, \mathbf{c}_{1}\right) \exp \left(-\gamma_{1}\left(t / T-c_{11}\right)\right) \gamma_{1}\right)^{\intercal} \\
&+\beta_{1} \frac{\partial \sigma_{t-1}^{2}}{\partial \boldsymbol{\theta}_{1}}
\end{align*}
and \begin{align*}\frac{\partial \sigma_{t}^{2}}{\partial \boldsymbol{\theta}_{2}}&=\frac{\partial}{\partial \boldsymbol{\theta}_{2}}\left\{\alpha_{0}+\alpha_{01} G_{1}\left(t / T ; \gamma_{1}, \mathbf{c}_{1}\right)+\alpha_{1} X_{t-1}^{2}+\beta_{1} \sigma_{t-1}^{2}\right\} \\
&=\left(1, X_{t-1}^{2}, \sigma_{t-1}^{2}\right)^{\intercal}+\beta_{1} \frac{\partial \sigma_{t-1}^{2}}{\partial \boldsymbol{\theta}_{2}}.\end{align*}

The partial derivatives corresponding to the alternative hypothesis are given in (\ref{eq:dft}). The recursions are infinite, but decay geometrically with $\beta_1.$ By using the analytic expressions of the derivatives, one has to truncate the geometric series at the first observed data point. This introduces an error in the expression for the score, especially for small $T$ and large $\beta_1$. However, since the test relies on an asymptotic distribution, this is not a concern in large samples. An alternative is to use numerically computed derivatives. Here, we shall use analytic derivatives.  

\subsection*{Calculation of the LM statistic by auxiliary regressions}
For the computation of the LM statistic, it is convenient to introduce the
matrix%
\begin{equation*}
\boldsymbol{\widehat{S}}=\left( 
\begin{array}{ccc}
\widehat{\boldsymbol{s}}^\intercal_{11} & \widehat{\boldsymbol{s}}^\intercal_{21} & \widehat{\boldsymbol{s}}^\intercal_{31} \\ 
\widehat{\boldsymbol{s}}^\intercal_{12} & \widehat{\boldsymbol{s}}^\intercal_{22} & \widehat{\boldsymbol{s}}^\intercal_{32} \\ 
\vdots & \vdots & \vdots \\ 
\widehat{\boldsymbol{s}}^\intercal_{1T} & \widehat{\boldsymbol{s}}^\intercal_{2T} & \widehat{\boldsymbol{s}}^\intercal_{3T}%
\end{array}%
\right)
\end{equation*}%
and the vector $\widehat{\boldsymbol{e}}=(\widehat{Z}_{1}^{2}-1, \widehat{Z}_{2}^{2}-1,\ldots, \widehat{Z}_{T}^{2}-1)^{\intercal}.$
The LM statistic can be written as
\begin{equation*}
LM=\frac{1}{2}\widehat{\boldsymbol{e}}^{\intercal }\boldsymbol{\widehat{S}}(\boldsymbol{\widehat{S}}^{\intercal}\boldsymbol{\widehat{S}})^{-1}\boldsymbol{\widehat{S}}^{\intercal}\widehat{\boldsymbol{e}}.
\end{equation*}
Noting that under normality,
\begin{equation*}
\mbox{plim}_{T\rightarrow \infty }\frac{\widehat{\boldsymbol{e}}^{\intercal}\widehat{\boldsymbol{e}}}{T}=2,
\end{equation*}%
an asymptotically equivalent statistic is%
\begin{equation*}
LM=T\frac{\widehat{\boldsymbol{e}}^{\intercal}\boldsymbol{\widehat{S}}(\boldsymbol{\widehat{S}}^{\intercal }\boldsymbol{\widehat{S}})^{-1}\boldsymbol{\widehat{S}}^{\intercal }\widehat{\boldsymbol{e}}}{\widehat{\boldsymbol{e}}^{\intercal}\widehat{\boldsymbol{e}}}=TR^{2},
\end{equation*}%
where $R^{2}$ is the $R^{2}$ from a regression of $\widehat{\boldsymbol{e}}$ on  $\boldsymbol{\widehat{S}}$. This form of the LM\ test in ARCH\ models was originally
suggested by \textcite{Engle82}. The derivatives in $\widehat{\boldsymbol{s}}_{3}$ are here approximated by the partial derivatives of the polynomial $\boldsymbol{\delta}^{\intercal} \boldsymbol{\tau}_{t}$ in (\ref{eq:unidLM2}). For notational brevity, define
\begin{equation*}
\widehat{\boldsymbol{r}}_{1t}:= (\widehat{\boldsymbol{s}}^\intercal_{1t}, \widehat{\boldsymbol{s}}^\intercal_{2t})^\intercal   
\end{equation*}
and
\begin{equation*}
 \widehat{\boldsymbol{r}}_{2t}:=  \frac{1}{\widehat{\sigma}^2_t}\left(\boldsymbol{\tau}_t + \widehat{\beta}_1\boldsymbol{\tau}_{t-1} + \widehat{\beta}_1^2\boldsymbol{\tau}_{t-2} + \cdots \right). \label{eq:r2t}
\end{equation*}

The following algorithm can be used to calculate the LM statistic.
\subsubsection*{LM test}
\begin{enumerate}
    \item Estimate the model under the null hypothesis. Save the standardized residuals $\widehat{Z}_t=X_t/\widehat{\sigma}_t.$ Construct the residual sum of squares $SSR_0=\sum_{t=1}^T\left(\widehat{Z}^2_t-1\right)^2.$
    \item Regress $\left(\widehat{Z}_t^2-1\right)$ on $\widehat{\boldsymbol{r}}_{1t}$ and $\widehat{\boldsymbol{r}}_{2t}$. Form the residual sum of squares $SSR_1$.
    \item Compute the test statistic $$LM=T\frac{SSR_0-SSR_1}{SSR_0}.$$
\end{enumerate}
Assuming that $Z_t$ is standard normal and that the order of the polynomial is three as in (\ref{eq:unidLM2}), the asymptotic distribution of the test statistic $LM$ is $\chi^2(3)$. 

The test statistic can be made robust against violations of distributional assumptions.
\textcite[Procedure 2.1]{wooldridge1990} shows that the robust version can be calculated as follows. 

\subsubsection*{Robust LM test}

\begin{enumerate}
    \item Estimate the model under the null hypothesis. Save the standardized residuals $\widehat{Z}_t=X_t/\widehat{\sigma}_t.$
    \item Regress $\widehat{\boldsymbol{r}}_{2t}$ component-wise on $\widehat{\boldsymbol{r}}_{1t}$ and compute the residual vector $\widehat{\boldsymbol{w}}_t=\left(\widehat{w}_{1t},\widehat{w}_{2t},\widehat{w}_{3t}\right)^{\intercal}$.
    \item Regress a constant $1$ on $\left(\widehat{Z}^2_t-1\right)\widehat{\boldsymbol{w}}_t.$ Compute the residual sum of squares $SSR$ and the test statistic $$LMr=T-SSR.$$
\end{enumerate}
The asymptotic distribution of the test statistic $LMr$ under the null hypothesis is again $\chi^2(3)$ if the order of the polynomial is three as in (\ref{eq:unidLM2}).

In theory, under the null, the elements of $\widehat{\boldsymbol{r}}_{1t}$ are orthogonal to $\widehat{Z}_{t}^{2}-1=X^2_t/\widehat{\sigma}^2_t - 1.$ However, due to the numerical optimization, this might not hold in practice. We therefore orthogonalize this term by regressing it onto $\widehat{\boldsymbol{r}}_{1t}$ and use the residuals from this regression as $\widehat{Z}^2_t$ in our tests. A similar stabilizing correction was used by \textcite{eitrheim1996} in a STAR context, and by \textcite{EngleNg93} in a GARCH framework. To improve the numerical accuracy of the estimation of the shape parameter $\gamma_{1}$, we apply the transformation $\gamma_{1}=\eta_{1}/(1-\eta_{1})^{-1}$ proposed by \textcite{ekner2013parameter} when estimating the ATV-GARCH model.
\newpage
\subsection*{B Simulation details and tables}

ML estimation is done numerically using the solver \texttt{Solnp} by \textcite{Ye1987}, implemented in the \texttt{R} package \texttt{Rsolnp} by \textcite{Rsolnp}. We give the solver the correct starting values since the aim is to investigate the properties of the tests rather than the numerical accuracy of the algorithm. We use a burn-in period of $200$ observations. The number of Monte Carlo replications is $5000$. To reduce the computational burden, we initialize the conditional variance by choosing a starting value as suggested in \textcite{FrancqZakoian2004}, rather than relying on a truncated recursive structure as in \textcite{BHK2003} and ABT. In the stationary null model case, we use the sample variance. When the null model is nonstationary as in testing an ATV-GARCH model with $L$ transitions against the same model with $L+1$ transitions, we use the square of the first observed value of the process.

The simulated rejection probabilities under the null hypothesis at the $1\%$, $5\%$ and $10\%$ significance levels are reported in Table \ref{tab:size}. The simulated powers at the $5\%$ and $10\%$ significance levels are reported in Table \ref{tab:power}.

\begin{table}
\centering
\caption{Simulated size.}
\begin{tabular}[t]{cc|cccccc}
\toprule
& & \multicolumn{2}{c}{LM} & & \multicolumn{2}{c}{Robust LM} & \\
\cmidrule(l{3pt}r{3pt}){3-5} \cmidrule(l{3pt}r{3pt}){6-8}\\
DGP & T & 1\% & 5\% & 10\% & 1\% & 5\% & 
10\%\\
\midrule
&& \multicolumn{6}{c}{\textbf{Panel 1}: 0 vs. 1 transition}\\
\cmidrule(l{3pt}r{3pt}){3-8} \\
1 & 1000 & 0.0268 & 0.0966 & 0.1750 & 0.0164 & 0.0696 & 0.1380\\
 & 2500 & 0.0160 & 0.0702 & 0.1244 & 0.0118 & 0.0588 & 0.1106\\
 & 5000 & 0.0120 & 0.0628 & 0.1176 & 0.0110 & 0.0550 & 0.1142\\
\midrule 
2 & 1000 & \cellcolor{gray!10}{0.0310} & \cellcolor{gray!10}{0.1060} & \cellcolor{gray!10}{0.1806} & \cellcolor{gray!10}{0.0212} & \cellcolor{gray!10}{0.0856} & \cellcolor{gray!10}{0.1574}\\
 & 2500 & \cellcolor{gray!10}{0.0176} & \cellcolor{gray!10}{0.0676} & \cellcolor{gray!10}{0.1292} & \cellcolor{gray!10}{0.0122} & \cellcolor{gray!10}{0.0604} & \cellcolor{gray!10}{0.1196}\\
 & 5000 & \cellcolor{gray!10}{0.0130} & \cellcolor{gray!10}{0.0658} & \cellcolor{gray!10}{0.1180} & \cellcolor{gray!10}{0.0120} & \cellcolor{gray!10}{0.0596} & \cellcolor{gray!10}{0.1116}\\
 \midrule
3 & 1000 & 0.0170 & 0.0648 & 0.1206 & 0.0132 & 0.0602 & 0.1176\\
 & 2500 & 0.0108 & 0.0516 & 0.1076 & 0.0104 & 0.0514 & 0.1030\\
 & 5000 & 0.0104 & 0.0530 & 0.1044 & 0.0112 & 0.0528 & 0.1014\\
\midrule
3t & 1000 & \cellcolor{gray!10}{0.0212} & \cellcolor{gray!10}{0.0686} & \cellcolor{gray!10}{0.1302} & \cellcolor{gray!10}{0.0114} & \cellcolor{gray!10}{0.0542} & \cellcolor{gray!10}{0.1154}\\
 & 2500 & \cellcolor{gray!10}{0.0152} & \cellcolor{gray!10}{0.0590} & \cellcolor{gray!10}{0.1128} & \cellcolor{gray!10}{0.0122} & \cellcolor{gray!10}{0.0586} & \cellcolor{gray!10}{0.1154}\\
 & 5000 & \cellcolor{gray!10}{0.0146} & \cellcolor{gray!10}{0.0524} & \cellcolor{gray!10}{0.0992} & \cellcolor{gray!10}{0.0100} & \cellcolor{gray!10}{0.0542} & \cellcolor{gray!10}{0.1018}\\
\midrule 
&& \multicolumn{6}{c}{\textbf{Panel 2}: 1 vs. 2 transitions}\\
\cmidrule(l{3pt}r{3pt}){3-8} \\
4 & 1000 & 0.0214 & 0.0852 & 0.1582 & 0.0182 & 0.0792 & 0.1528\\
 & 2500 & 0.0186 & 0.0740 & 0.1330 & 0.0122 & 0.0676 & 0.1278\\
 & 5000 & 0.0168 & 0.0678 & 0.1234 & 0.0134 & 0.0634 & 0.1172\\
\midrule 
5 & 1000 & \cellcolor{gray!10}{0.0178} & \cellcolor{gray!10}{0.0676} & \cellcolor{gray!10}{0.1220} & \cellcolor{gray!10}{0.0138} & \cellcolor{gray!10}{0.0594} & \cellcolor{gray!10}{0.1130}\\
 & 2500 & \cellcolor{gray!10}{0.0136} & \cellcolor{gray!10}{0.0614} & \cellcolor{gray!10}{0.1078} & \cellcolor{gray!10}{0.0122} & \cellcolor{gray!10}{0.0546} & \cellcolor{gray!10}{0.1080}\\
 & 5000 & \cellcolor{gray!10}{0.0134} & \cellcolor{gray!10}{0.0594} & \cellcolor{gray!10}{0.1136} & \cellcolor{gray!10}{0.0106} & \cellcolor{gray!10}{0.0558} & \cellcolor{gray!10}{0.1080}\\
\midrule
6 & 1000 & 0.0120 & 0.0526 & 0.0990 & 0.0082 & 0.0474 & 0.0902\\
 & 2500 & 0.0112 & 0.0520 & 0.0952 & 0.0094 & 0.0450 & 0.0922\\
 & 5000 & 0.0092 & 0.0468 & 0.0932 & 0.0080 & 0.0470 & 0.0880\\
 \midrule
7 & 1000 & \cellcolor{gray!10}{0.0216} & \cellcolor{gray!10}{0.0782} & \cellcolor{gray!10}{0.1404} & \cellcolor{gray!10}{0.0176} & \cellcolor{gray!10}{0.0690} & \cellcolor{gray!10}{0.1290}\\
 & 2500 & \cellcolor{gray!10}{0.0148} & \cellcolor{gray!10}{0.0686} & \cellcolor{gray!10}{0.1262} & \cellcolor{gray!10}{0.0118} & \cellcolor{gray!10}{0.0610} & \cellcolor{gray!10}{0.1202}\\
 & 5000 & \cellcolor{gray!10}{0.0130} & \cellcolor{gray!10}{0.0642} & \cellcolor{gray!10}{0.1246} & \cellcolor{gray!10}{0.0122} & \cellcolor{gray!10}{0.0570} & \cellcolor{gray!10}{0.1172}\\
 \midrule
8 & 1000 & 0.0148 & 0.0618 & 0.1060 & 0.0096 & 0.0504 & 0.1024\\
 & 2500 & 0.0114 & 0.0544 & 0.1072 & 0.0104 & 0.0514 & 0.1024\\
 & 5000 & 0.0112 & 0.0540 & 0.1086 & 0.0098 & 0.0514 & 0.1028\\
 \midrule
9 & 1000 & \cellcolor{gray!10}{0.0116} & \cellcolor{gray!10}{0.0478} & \cellcolor{gray!10}{0.0900} & \cellcolor{gray!10}{0.0088} & \cellcolor{gray!10}{0.0452} & \cellcolor{gray!10}{0.0822}\\
 & 2500 & \cellcolor{gray!10}{0.0102} & \cellcolor{gray!10}{0.0410} & \cellcolor{gray!10}{0.0766} & \cellcolor{gray!10}{0.0074} & \cellcolor{gray!10}{0.0366} & \cellcolor{gray!10}{0.0752}\\
 & 5000 & \cellcolor{gray!10}{0.0092} & \cellcolor{gray!10}{0.0430} & \cellcolor{gray!10}{0.0852} & \cellcolor{gray!10}{0.0094} & \cellcolor{gray!10}{0.0416} & \cellcolor{gray!10}{0.0788}\\
\bottomrule
\end{tabular}
\label{tab:size}
\end{table}

\begin{table}
\centering
\caption{Simulated power.}
\begin{tabular}[t]{cc|cccccc}

\toprule
& & \multicolumn{2}{c}{LM} & & \multicolumn{2}{c}{Robust LM} & \\
\cmidrule(l{3pt}r{3pt}){3-5} \cmidrule(l{3pt}r{3pt}){6-8}\\
DGP & T & 1\% & 5\% & 10\% & 1\% & 5\% & 
10\%\\
\midrule
&& \multicolumn{6}{c}{\textbf{Panel 1}: 0 vs. 1 transition}\\
\cmidrule(l{3pt}r{3pt}){3-8} \\
4 & 1000 & 0.9852 & 0.9990 & 0.9996 & 0.9116 & 0.9926 & 0.9982\\
& 2500 & 0.9998 & 1.0000 & 1.0000 & 0.9982 & 1.0000 & 1.0000\\
& 5000 & 1.0000 & 1.0000 & 1.0000 &  1.0000 & 1.0000 & 1.0000\\
\midrule
5 & 1000 & \cellcolor{gray!10}{0.7720} & \cellcolor{gray!10}{0.9406} & \cellcolor{gray!10}{0.9754} & \cellcolor{gray!10}{0.6566} & \cellcolor{gray!10}{0.9024} & \cellcolor{gray!10}{0.9634}\\
& 2500 & \cellcolor{gray!10}{0.9794} & \cellcolor{gray!10}{0.9998} & \cellcolor{gray!10}{1.0000} & \cellcolor{gray!10}{0.9676} & \cellcolor{gray!10}{0.9978} & \cellcolor{gray!10}{1.0000}\\
& 5000 & \cellcolor{gray!10}{0.9982} & \cellcolor{gray!10}{1.0000} & \cellcolor{gray!10}{1.0000} & \cellcolor{gray!10}{0.9972} & \cellcolor{gray!10}{0.9998} & \cellcolor{gray!10}{1.0000}\\
\midrule
6 & 1000 & 0.3144 & 0.5656 & 0.6978 & 0.2584 & 0.5260 & 0.6652\\
& 2500 & 0.7970 & 0.9376 & 0.9712 & 0.7654 & 0.9322 & 0.9676\\
& 5000 & 0.9920 & 0.9990 & 0.9998 & 0.9898 & 0.9990 & 0.9998\\
\midrule
7 & 1000 & \cellcolor{gray!10}{0.9438} & \cellcolor{gray!10}{0.9952} & \cellcolor{gray!10}{0.9978} & \cellcolor{gray!10}{0.8638} & \cellcolor{gray!10}{0.9828} & \cellcolor{gray!10}{0.9960}\\
& 2500 & \cellcolor{gray!10}{0.9986} & \cellcolor{gray!10}{1.0000} & \cellcolor{gray!10}{1.0000} & \cellcolor{gray!10}{0.9956} & \cellcolor{gray!10}{1.0000} & \cellcolor{gray!10}{1.0000}\\
& 5000 & \cellcolor{gray!10}{0.9998} & \cellcolor{gray!10}{1.0000} & \cellcolor{gray!10}{1.0000} & \cellcolor{gray!10}{1.0000} & \cellcolor{gray!10}{1.0000} & \cellcolor{gray!10}{1.0000}\\
\midrule
8 & 1000 & 0.4882 & 0.7468 & 0.8448 & 0.4196 & 0.7106 & 0.8254\\
& 2500 & 0.9324 & 0.9886 & 0.9956 & 0.9172 & 0.9858 & 0.9960\\
& 5000 & 0.9980 & 0.9998 & 1.0000 & 0.9970 & 0.9998 & 0.9998\\
\midrule
9 & 1000 & \cellcolor{gray!10}{0.1528} & \cellcolor{gray!10}{0.3430} & \cellcolor{gray!10}{0.4788} & \cellcolor{gray!10}{0.1244} & \cellcolor{gray!10}{0.3202} & \cellcolor{gray!10}{0.4558}\\
& 2500 & \cellcolor{gray!10}{0.4584} & \cellcolor{gray!10}{0.7034} & \cellcolor{gray!10}{0.8134} & \cellcolor{gray!10}{0.4376} & \cellcolor{gray!10}{0.6998} & \cellcolor{gray!10}{0.8138}\\
& 5000 & \cellcolor{gray!10}{0.8834} & \cellcolor{gray!10}{0.9686} & \cellcolor{gray!10}{0.9854} & \cellcolor{gray!10}{0.8758} & \cellcolor{gray!10}{0.9670} & \cellcolor{gray!10}{0.9848}\\
\midrule
10 & 1000 & 0.7562 & 0.9416 & 0.9788 & 0.7198 & 0.9278 & 0.9734\\
& 2500 & 0.9700 & 0.9968 & 1.0000 & 0.9560 & 0.9938 & 0.9990\\
& 5000 & 0.9964 & 1.0000 & 1.0000 & 0.9950 & 0.9998 & 1.0000\\
\midrule
11 & 1000 & \cellcolor{gray!10}{0.8728} & \cellcolor{gray!10}{0.9756} & \cellcolor{gray!10}{0.9918} & \cellcolor{gray!10}{0.8022} & \cellcolor{gray!10}{0.9622} & \cellcolor{gray!10}{0.9884}\\
& 2500 & \cellcolor{gray!10}{0.9944} & \cellcolor{gray!10}{1.0000} & \cellcolor{gray!10}{1.0000} & \cellcolor{gray!10}{0.9910} & \cellcolor{gray!10}{0.9998} & \cellcolor{gray!10}{1.0000}\\
& 5000 & \cellcolor{gray!10}{0.9998} & \cellcolor{gray!10}{1.0000} & \cellcolor{gray!10}{1.0000} & \cellcolor{gray!10}{0.9996} & \cellcolor{gray!10}{1.0000} & \cellcolor{gray!10}{1.0000}\\
\midrule
&& \multicolumn{6}{c}{\textbf{Panel 2}: 1 vs. 2 transitions}\\
\cmidrule(l{3pt}r{3pt}){3-8} \\
10 & 1000 & 0.5908 & 0.8182 & 0.8994 & 0.6542 & 0.8496 & 0.9138\\
& 2500 & 0.9626 & 0.9850 & 0.9892 & 0.9602 & 0.9842 & 0.9886\\
& 5000 & 0.9780 & 0.9856 & 0.9900 & 0.9770 & 0.9858 & 0.9898\\
\midrule
11 & 1000 & \cellcolor{gray!10}{0.0224} & \cellcolor{gray!10}{0.0848} & \cellcolor{gray!10}{0.1536} & \cellcolor{gray!10}{0.0160} & \cellcolor{gray!10}{0.0680} & \cellcolor{gray!10}{0.1364}\\
& 2500 & \cellcolor{gray!10}{0.0218} & \cellcolor{gray!10}{0.0882} & \cellcolor{gray!10}{0.1618} & \cellcolor{gray!10}{0.0180} & \cellcolor{gray!10}{0.0744} & \cellcolor{gray!10}{0.1444}\\
& 5000 & \cellcolor{gray!10}{0.0300} & \cellcolor{gray!10}{0.1020} & \cellcolor{gray!10}{0.1802} & \cellcolor{gray!10}{0.0236} & \cellcolor{gray!10}{0.0962} & \cellcolor{gray!10}{0.1684}\\
\bottomrule
\end{tabular}
\label{tab:power}
\end{table}

\end{document}